\documentclass[twocolumn,amsmath,amssymb]{revtex4} 
\usepackage{graphicx}
\usepackage{dcolumn}
\usepackage{bm}
\begin{document} 
 
\title{Misoriented bilayer graphene in a magnetic field: Optical 
transitions at commensurate angles} 
\author{Vadim M. Apalkov} 
\affiliation{Department of Physics and Astronomy, Georgia State University, 
Atlanta, Georgia 30303, USA} 
\author{Tapash Chakraborty$^\ddag$} 
\affiliation{Department of Physics and Astronomy, 
University of Manitoba, Winnipeg, Canada R3T 2N2} 
 
\date{\today} 
\begin{abstract} 
Misoriented bilayer graphene with commensurate angles shows unique magneto-optical  
properties. The optical absorption spectra of such a system strongly depend on the  
angle of rotation. For a general commensurate twist angle the absorption spectra  
has a simple single-peak structure. However, our studies indicate that there  
are special angles at which the absorption spectra of the rotated bilayer  
exhibit well developed multi-peak structures. These angles correspond to even  
symmetry of the rotated graphene with respect to the sublattice exchange. Magneto-%
spectroscopy can therefore be a potentially useful scheme to determine the twist 
angles. 
\end{abstract} 
\maketitle 
 
Graphene, mechanically exfoliated from graphite \cite{novo} display truly 
remarkable electronic properties \cite{review}, and holds an immense potential 
to become a key ingradient for a new generation of electronic devices. The 
dynamics of electrons in a single sheet of graphene, a hexagonal honeycombed 
lattice of carbon atoms is that of massless Dirac fermions with linear 
dispersion, chiral eigenstates, valley degeneracy, and unusual Landau levels 
in an external magnetic field \cite{review}. Bilayer graphene, on the other 
hand, show quadratic dispersion \cite{mccann} and the charge carriers there 
are characterized as massive chiral fermions. Interestingly, epitaxial graphene 
\cite{epitaxial}, which is thermally grown on the C face of the SiC substrate, 
as well as graphene grown by chemical vapor deposition (CVD) \cite{reina}, are 
multilayer films and yet, quite surprisingly display behavior similar to that of a 
single layer graphene \cite{walt_09}. These systems are known to have a high  
degree of rotational misalignments \cite{hass}. Theoretical studies of  
turbostratic bilayer graphene \cite{latil,turbo,lopes07,mele10} have indicated 
that in this case the interlayer coupling is suppressed and the systems can be 
roughly considered as two decoupled layers of graphene. At the same time due to 
the modulated nature \cite{lopes07} of the interlayer transfer integral, these 
systems show quite rich low-energy physics, which strongly depends on the nature 
of commensurate stacking faults \cite{mele10}. 
 
In a rotated bilayer graphene the rotational stacking fault is determined 
by an angle $\theta$ of rotation of one layer relative to the  
other [see Fig.~1(a)]. Each layer consists of two sublattices,  
A and B, and is characterized by two primitive translational lattice  
vectors: $\vec{a}=a(1,0)$ and $\vec{b}=a(-1/2,\sqrt{3}/2)$, where $a=0.246$  
nm is the lattice constant. The commensurate rotation in a twisted bilayer  
is defined by the condition \cite{lopes07}, $\vec{T}^{}_{mk}=m\vec{a}+k\vec{b} 
=\vec{T}^{}_{m^{\prime} k^{\prime}}= m^{\prime}\vec{a}^{\prime}+k^{\prime} 
\vec{b}^{\prime}$, where $\vec{a}^{\prime}$ and $\vec{b}^{\prime}$ are  
given by the rotation of the primitive vectors $\vec{a}$ and $\vec{b}$ by  
$\theta$. The angles corresponding to the commensurate stacking fault are  
determined from: $\cos\theta=(3q^2-p^2)/(3q^2+p^2)$, where $q>p>0$ are  
integers. There are two types of commensurate rotations that are distinguished  
by their symmetry, even or odd, with respect to the sublattice exchange  
\cite{mele10}. For the even commensurate stacking fault both A and B sublattice  
sites of the two layers are coincident at some point, while for  
the odd stacking fault only A sublattice sites of two layers are coincident.  
The AA-stacking and Bernal stacking correspond to even and odd stacking  
faults with angles $\theta=0$ and $60^\circ$, respectively. 
 
\begin{figure} 
\begin{center}\includegraphics[width=6.5cm]{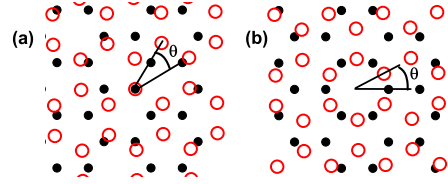}\end{center} 
\vspace*{-0.8cm} 
\caption{ 
Misoriented graphene bilayer with angle of rotation $\theta$, shown 
schematically in real space (a) and reciprocal space (b). The black  
solid dots and red open dots are the (a) atomic positions in real space,  
and (b) $\vec{K}$ and $\vec{K}^{\prime}$ points in the reciprocal space in 
different layers. 
} 
\label{figone} 
\end{figure} 
 
The reciprocal lattice of a graphene layer consist of $K$ and $K^{\prime}$  
sets of points: $\vec{K}+\vec{G}^{}_{m,k}$, $\vec{K}^{\prime }+\vec{G}^{}_{m,k}$,  
where $\vec{G}^{}_{m,k}=m\vec{G}^{}_1+k\vec{G}^{}_2$, $m$ and $k$ are  
integers, $\vec{G}^{}_1=2\pi/a(1,1/\sqrt{3})$ and $\vec{G}^{}_2=2\pi/a(0, 2 
/\sqrt{3})$ are primitive reciprocal lattice vectors, and $\vec{K}=2\pi/a(1/3,  
1/\sqrt{3})$, $\vec{K}^{\prime}=2\pi/a(2/3,0)$. For a rotated  
layer the reciprocal lattice is rotated by an angle $\theta$ around the  
origin [see Fig.~1(b)]. Then the even commensurate stacking fault corresponds  
to the twist angles at which the $K$ points of reciprocal lattice of two  
layers are coincident \cite{mele10}, i.e., $\vec{K} + \vec{G}^{}_{m,k}=\vec{K}(\theta)+ 
\vec{G}^{}_{m^{\prime},k^{\prime}} (\theta)$ at some values of $k$, $m$,  
$k^{\prime}$, and $m^{\prime}$. For odd  stacking fault the $K$ 
and $K^{\prime}$ points are coincident, i.e., $\vec{K} + \vec{G}^{}_{m,k} 
= \vec{K}^{\prime}(\theta)+\vec{G}^{}_{m^{\prime},k^{\prime}}(\theta)$ \cite{mele10}. 
 
In the case of the commensurate stacking fault the interlayer coupling  
becomes a periodically modulated function of position, which results in  
the interlayer coupling determined only by the coincident points of the  
reciprocal lattice that are found from $\vec{K}+\vec{G}^{}_{m,k}=\vec{K} 
(\theta)+\vec{G}^{}_{m^{\prime}, k^{\prime}}(\theta)$ or $\vec{K}+ 
\vec{G}^{}_{m,k} = \vec{K}^{\prime}(\theta)+\vec{G}^{}_{m^{\prime},k^{\prime}}  
(\theta)$. Therefore the interlayer coupling is characterized by the  
Fourier transform of the interlayer potential, $t(k)$, taken at points  
$\vec{K} + \vec{G}^{}_{m,k}$: $t^{}_{m,k} = t(\vec{K} + \vec{G}^{}_{m,k})$. 
As a result, the effective low energy Hamiltonian of the twisted  
layer at commensurate condition takes the form \cite{mele10} 
\begin{equation} 
{\cal H}^{}_{even} = \left( 
\begin{array}{cccc} 
0 & v^{}_F \hat{p}^{}_{1,-} & t^{}_{\theta} e^{i \phi/2} & 0 \\ 
v^{}_F \hat{p}^{}_{1,+} & 0 & 0 & t^{}_{\theta} e^{-i \phi /2} \\ 
t_{\theta}^{+} e^{i \phi/2} & 0 & 0 & v^{}_F \hat{p}^{}_{2,-} \\ 
0 & t_{\theta}^{+} e^{i \phi/2} & v^{}_F \hat{p}^{}_{2,+} & 0 
\end{array} 
\right) 
\label{Heven} 
\end{equation} 
\begin{equation} 
{\cal H}_{odd} = \left( 
\begin{array}{cccc} 
0 & v^{}_F\hat{p}^{}_{1,-} & t^{}_{\theta} & 0 \\ 
v^{}_F\hat{p}^{}_{1,+} & 0 & 0 & 0 \\ 
t_{\theta }^{+} & 0 & 0 & v^{}_F \hat{p}^{}_{2,+} \\ 
0 & 0 & v^{}_F \hat{p}^{}_{2,-} & 0 
\end{array} 
\right) 
\label{Hodd} 
\end{equation} 
where $t^{}_{\theta}=t^{}_{m,k} e^{i\theta}$, $p^{}_{\alpha,\pm}=p^{}_{\alpha,x} 
\pm i p^{}_{i,y}$ is the electron momentum operator for layer $\alpha=1,2$,  
$v^{}_F \approx 10^6$ m/s. The phase angle $\phi$ is determined from  
$\phi=2\pi/3(m-k)$, which can have possible values 0 or $\pm 2\pi/3$. For example, 
for $\theta=38.2^\circ$ the phase angle is $\phi=2\pi/3$. 
 
For the odd-twisted bilayer the Hamiltonian for all twist angles  
is identical to the Hamiltonian of a bilayer graphene with Bernal stacking,  
but with suppressed interlayer coupling. We found new and unique features for  
the even-twisted bilayer. While at $\phi=0$ the Hamiltonian of a twisted  
bilayer is similar to that of a bilayer with AA stacking, at $\phi=\pm 2\pi/3$   
the Hamiltonian of a twisted graphene becomes unique. We show below that this  
Hamiltonian exhibits interesting features in positions of the Landau levels (LLs)  
and in magneto-spectroscopy of the rotated bilayer. 
 
Introducing a magnetic field in the Hamiltonians (\ref{Heven})-(\ref{Hodd}) 
by replacing the momentum operator $\vec{p}$ with $\vec{p} + e\vec{A}/c$,  
where $\vec{A}$ is the vector potential, we obtain the LL  
energies of a twisted bilayer. For odd and even bilayers the interlayer  
tunneling introduces different coupling structures. For the odd bilayer  
the LLs of individual layers belonging to different LL indices are coupled,  
while for the even bilayer the LLs of same indices are coupled. 
 
For the odd bilayer the LL energy spectrum is similar to that of a bilayer 
graphene with Bernal stacking 
\begin{equation*} 
\varepsilon = s^{}_0 \varepsilon^{}_0 \sqrt{2n +1+ \frac{t_{m,k}^2}{2 
\varepsilon_0^2}+s^{}_1 \sqrt{\left(1+\frac{t_{m,k}^2}{2\varepsilon_0^2} 
\right )^2+2n\frac{t^{}_{m,k}}{\varepsilon^{}_0}}}, 
\label{LL_odd} 
\end{equation*} 
where $n=0,1,2,\ldots$, $s^{}_0=\pm 1$ corresponds to the conduction and  
valence bands, respectively, and $s^{}_1=\pm 1$ determines the splitting of  
levels due to interlayer coupling. Here $\varepsilon^{}_0=\hbar v^{}_F/\ell^{}_0$  
and $\ell^{}_0=\sqrt{c\hbar/eB}$. For each $n$ there are two  
particle-like and two hole-like LLs. 
 
\begin{figure} 
\begin{center}\includegraphics[width=7.0cm]{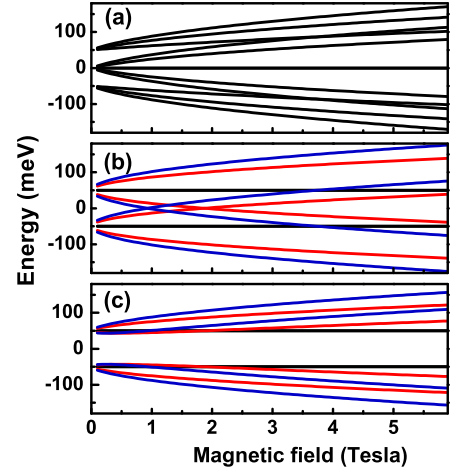}\end{center} 
\vspace*{-0.8cm} 
\caption{ 
A few lowest LLs of three different types of misoriented  
bilayer versus the magnetic field: (a) the odd-twisted bilayer;  
(b) the even-twisted bilayer with phase angle $\phi=0$; (c) the even-twisted  
bilayer with phase angle $\phi=2\pi/3$. The interlayer coupling  
is set to 50 meV. Red and blue lines in (b) and (c) correspond to the 
LL index $n=1,2$ respectively. 
} 
\label{figtwo} 
\end{figure} 
 
For the even-twisted bilayer the interlayer tunneling couples the LLs 
of the two layers with same LL indices, resulting in a splitting  
of the originally degenerate levels. This splitting depends on $\phi$.  
The energy spectrum of a twisted graphene in this case is 
\begin{equation} 
\varepsilon=s^{}_0\varepsilon^{}_0\sqrt{2n+\frac{t_{m,k}^2}{\varepsilon_0^2} 
+2s^{}_1\sqrt{2n}\frac{t^{}_{m,k}}{\varepsilon^{}_0}\cos(\phi/2)}, 
\label{LL_even} 
\end{equation} 
where $n=0,1,2,\ldots$, $s^{}_0=\pm 1$, $s^{}_1=\pm 1$. If $\phi = 0$  
then $\varepsilon=\pm\sqrt{2n}\varepsilon^{}_0\pm t^{}_{m,k}$, which  
is a simple splitting of degenerate LLs of the two graphene layers.  
This energy spectrum is similar to that of AA stacked bilayer.  
If $\phi=\pm 2\pi/3$ then $\varepsilon=\pm\sqrt{2n\varepsilon_0^2+t_{m,k}^2 
\pm\sqrt{2n}\varepsilon^{}_0 t^{}_{m,k}}$. Here the energy splitting of  
degenerate LLs of two layers is less than the corresponding  
splitting for $\phi=0$. 
 
In Fig.~2, the magnetic field dependence of LLs for different types  
of twisted bilayer are shown for interlayer  
coupling strength of 50 meV. For the odd graphene bilayer [Fig.~2(a)] the  
energy spectra is similar to that of a bilayer with Bernal stacking. There  
is one level with exactly zero energy, which does not show any magnetic  
field dependence. For the even-twisted bilayer there are two levels 
without any magnetic field dispersion [Fig.~2(b)]. These levels correspond  
to the LL index $n=0$ where the energy of the LLs for all values of  
$\phi$ is $\varepsilon=\pm t^{}_{m,k}$. For $n>0$ the LL spectrum  
depends on the value of $\phi $. The LL spectrum of even-twisted  
graphene with $\phi=0$ [Fig.~2(b)] is similar to that of a bilayer  
with AA stacking. The LL spectrum of a graphene bilayer with $\phi=\pm  
2\pi/3$ [Fig.~2(c)] shows an unique feature: there is a clear gap-like  
behavior in the spectrum with no LLs within the interval $(\sqrt{3}/2) t^{}_{m,k} 
>\varepsilon>-(\sqrt{3}/2) t^{}_{m,k}$ for all values of the magnetic field. 
 
\begin{figure} 
\begin{center}\includegraphics[width=6.5cm]{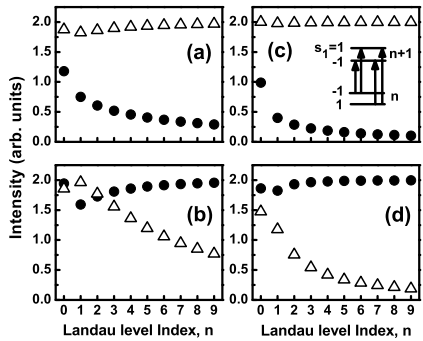}\end{center} 
\vspace*{-0.8cm} 
\caption{ 
Intensity of optical transitions (absorption) between the valence  
($s^{}_{0,{\bf i}}=-1$) and conduction ($s^{}_{0,{\bf f}}=1$) bands  
shown for even-twisted bilayer with phase angle $\phi=2\pi/3$  
as a function of the LL index. The magnetic field is 0.5 Tesla  
for (a) and (b), and 2 Tesla for (c) and (d). The optical transitions  
are between the initial states with $s^{}_{1,{\bf i}}=1$ [(a) and (c)]  
or $s^{}_{1,{\bf i}}=-1$ [(b) and (d)] and the final states with  
$s^{}_{1,{\bf f}}=1$ (dots) or $s^{}_{1,{\bf f}}=-1$ (triangles).  
Inset: schematic illustration of optical transitions from the initial  
valence band states with LL index $n$ and different values of  
$s^{}_1$ to the final conduction band states with LL index $n+1$  
and different values of $s^{}_1$. 
} 
\label{figthree} 
\end{figure} 
 
The LL wavefunctions of the twisted bilayer  
also depend on the type of commensurate stacking fault. Below we consider  
in detail only the case of even bilayer, which shows new and interesting  
features. The wavefunctions for the even-twisted bilayer  
corresponding to the LLs (\ref{LL_even}) are 
\begin{equation} 
\Psi^{}_{n,s^{}_0,s^{}_1}=C^{}_n\left( 
\begin{array}{c} 
 s^{}_0s^{}_1 f^{}_n e^{i\theta /2}e^{i\beta}\psi^{}_{n-1} \\ 
 -i s^{}_1 e^{i\theta /2}\psi^{}_{n} \\ 
 f^{}_n e^{- i\theta /2}\psi^{}_{n-1}  \\ 
 - i s^{}_0 e^{- i\theta /2}e^{i\beta}\psi^{}_{n} 
\end{array} 
\right) 
\label{f1} 
\end{equation} 
where $\psi^{}_n$ is the wavefunction of the conventional LL with index $n$,  
$f^{}_n=1, C^{}_n = 1/2$ if $n>0$ and $f^{}_n=0, C^{}_n = 1/\sqrt{2}$ if $n=0$. 
The phase $\beta$ is defined as 
\begin{equation*} 
\beta=\arcsin\left[\frac{s^{}_1t^{}_{m,k}\sin(\phi/2)}{ 
\sqrt{2n \varepsilon_0^2 + t_{m,k}^2 
+2 s^{}_1 \sqrt{2n}t^{}_{m,k}\varepsilon^{}_0 \cos(\phi/2)}} 
 \right], 
\end{equation*} 
which is nonzero only for nonzero values of $\phi$. This illustrates the 
sensitivity of the LL functions of a twisted bilayer to the twist angle. 
 
Armed with the wavefunctions (\ref{f1}) we are now ready to evaluate the  
strength of the electron-electron interaction within a single LL. This  
interaction is characterized by the Haldane pseudopotentials \cite{haldane}  
and is responsible for formation of the fraction quantum Hall effect (FQHE) states  
\cite{fqhe_book}. Since the interlayer coupling in Eq.~(\ref{f1}) affects only  
the phases of the wavefunction components, the pseudopotentials for the  
wavefunctions (\ref{f1}) are determined only by the LL index $n$  
and does not depend on the interlayer coupling. The pseudopotentials for  
the even-twisted bilayer are identical to those of individual  
graphene layers. Hence the strength of the FQHE in the even-twisted bilayer  
is the same as in an isolated graphene layer \cite{vadim_fqhe}. Therefore, as  
far as the FQHE is concerned, the even-twisted bilayer can be considered as two  
{\em decoupled} graphene layers for any twist angle \cite{walt_10}. Since the  
odd-twisted bilayer is similar to the bilayer graphene with Bernal  
stacking, the FQHE in an odd bilayer is similar to a bilayer graphene with  
Bernal stacking \cite{bilayer}. 
 
The phase $\beta$ in Eq.~(\ref{f1}) depends on both the twist angle (through 
$\phi$) and the LL index $n$. Although this phase does not influence the  
interaction properties within a single LL, importantly, it can  
modify the optical transitions between different LLs. For light  
polarized along the $x$ axis, the operator $\hat{M}$ of optical transitions is  
\begin{equation} 
\hat{M} \propto  \left( 
\begin{array}{cc} 
\sigma^{}_x   &  0  \\ 
0          & \sigma^{}_x 
\end{array} 
  \right), 
\end{equation} 
where $\sigma^{}_x $ is the Pauli matrix. Then the selection rule for the  
optical transitions between the initial state $\bf i$, and the final state  
$\bf f$, is the same as for a single graphene layer, i.e., $n^{}_{\bf f}  
= n^{}_{\bf i} +\pm 1$. We consider only the optical transitions to  
higher excited state, i.e., $n^{}_{\bf f} = n^{}_{\bf i}+1$, as those  
have higher frequencies and are perhaps easier to observe.  
The intensity of the corresponding optical transitions is 
$$I_{\bf{if}} = I^{}_0 C_{n^{}_{\bf i}}^2 C_{n_{\bf f}}^2\left|s^{}_{0,{\bf i}}  
s^{}_{0,{\bf f}} s^{}_{1,{\bf i}} s^{}_{1,{\bf f}}+e^{-(\beta^{}_{\bf i} 
+\beta^{}_{\bf f})}\right|^2.$$ 
For the even-twisted bilayer with $\phi =0$ the phases $\beta^{}_{\bf i}$   
and $\beta^{}_{\bf f}$ are zero. Then the intraband optical transitions  
(the same sign of $s^{}_0$) are allowed only between the states with the same  
sign of $s^{}_1$, while the interband optical transitions are allowed between  
the states with opposite signs of $s^{}_1$. All transitions have the same  
intensity. Therefore for a given LL index $n^{}_{\bf i}$ the optical absorption  
consists of a {\em single line}. 
 
\begin{figure} 
\begin{center}\includegraphics[width=6.5cm]{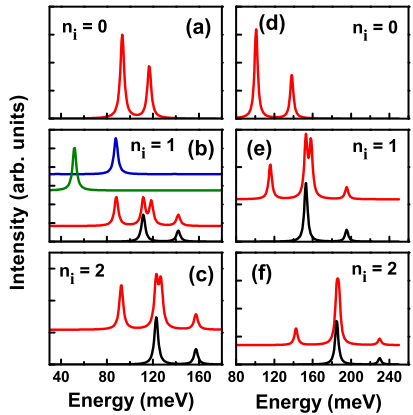}\end{center} 
\vspace*{-0.8cm} 
\caption{ 
Intensity of optical transitions (absorption) between the valence  
($s^{}_{0,{\bf i}}=-1$) and the conduction ($s^{}_{0,{\bf f}}=1$) bands  
is shown for the even-twisted bilayer with a phase angle  
$\phi = 2\pi/3$ as a function of the transition energy for different 
LL index $n_{\bf i}$ of the initial state. The magnetic field is 0.5 Tesla  
(left panel) and 2 Tesla (right panel). The black lines correspond to  
transitions from the initial state with $s^{}_1=1$ only, while the red  
lines correspond to transitions from all initial states ($s^{}_1=1$ and $-1$)  
with a given LL index $n$. For comparison, the optical absorption spectra  
of odd bilayer (blue line) and even bilayer with $\phi=0$  
(green line) are shown in panel (b) for $n_{\bf i}=1$.  
} 
\label{figfour} 
\end{figure} 
 
A different situation occurs for the even-twisted bilayer with nonzero  
values of $\phi$. In this case both $\beta^{}_{\bf i}$ and $\beta^{}_{\bf f}$  
are nonzero and depend on the LL index $n$. We consider below only the  
interband optical transitions, i.e., $s^{}_{0,{\bf i}}=-1$ in the initial  
state (valence band) and $s^{}_{0,{\bf f}}=1$ in the final state (conduction  
band). For each initial state there are two transitions to the final states  
with $s_{1,{\bf f}}=1$ and $-1$. In Fig.~3 the relative intensity of these  
transitions are shown for different initial LLs and for different magnetic  
fields. For each LL index the initial and final states are characterized by  
the parameter $s^{}_1$. The four possible optical transitions at a given LL  
index are shown schematically as inset in Fig.~3(c). In Fig.~3(a,c), the  
transitions from the states with $s^{}_{1,{\bf i}}=1$ are shown, while in  
Fig.~3(b,d) the initial state has $s^{}_{1,{\bf i}}=-1$. The triangles and  
dots correspond to the final states with $s^{}_{1,{\bf f}}=-1$ and $1$,  
respectively. The general tendency illustrated in Fig.~3 is the existence  
of strong transitions to {\em both final states}. At a small magnetic field  
and small LL index, these transitions have comparable intensities [see  
Fig.~3(a,b) where the magnetic field is 0.5 Tesla]. With increasing LL index  
the transition to one of the states is suppressed and the system becomes  
similar to the case of $\phi=0$. The regions of magnetic fields and LL  
indices for which the optical transitions have comparable intensities are  
determined by the strength of inter-layer transfer integral, $t_{m,k}$. The  
strongest coupling should be expected for small values of $m$ and $k$, e.g.,  
for the twist angle $38.2^0$.  
 
The unique features of optical transitions in the even-twisted bilayer with  
$\phi\neq0$ are clearly visible in the optical absorption spectra. In Fig.~4,  
we show the absorption spectra from a given LL of the valence band  
of a twisted bilayer with interlayer coupling of 50 meV. At a given LL  
index the absorption spectra shown by black solid lines correspond to  
transition from the ground state with $s^{}_{1,{\bf i}}=1$ [see the inset in  
Fig.~3(c)], i.e., only the state with $s^{}_{1,{\bf i}}=1$ is occupied. The  
absorption spectra from all the initial states ($s^{}_{1,{\bf i}}=1$ and $-1$)  
with a given LL index are shown by red lines. For the LL index  
$n=0$ there is only one initial state in the valence band. A 2 meV broadening  
of the optical lines is introduced in Fig.~4. The absorption spectra clearly  
show a multi-peak structure which is more pronounced at a small magnetic field  
[0.5 T in Fig.~4(a-c)] and at small LL index. With increasing magnetic  
field [see Fig.~4(d-f)] or increasing LL index only one peak in  
the absorption spectra survives, which is consistent with the results shown in  
Fig.~3. Such a behavior of the absorption spectra of even-twisted bilayer with  
$\phi\ne0$ is totally different from the odd-twisted graphene or even-twisted 
with $\phi=0$. It is easy to show that for the odd-twisted graphene, which is  
similar to bilayer graphene with Bernal stacking, only one strong optical  
transition exists for each LL. Therefore for both odd bilayer and  
even bilayer with $\phi=0$ the optical spectra for each LL consist of a single  
line [Fig.~4(b), $n_{\bf i}=1$]. Figure~4 clearly shows the fingerprints of  
phase angle $\phi$ in the magneto-optics of commensurate twisted graphene.  
 
In conclusion, magneto-optical properties of twisted graphene bilayer  
show strong dependence on the twist angle. At twist angles corresponding  
to odd bilayer and even bilayer with zero phase angle, $\phi =0$, the absorption  
spectra from a given LL consist of a single line, while the optical  
spectrum of the even bilayer with nonzero angle $\phi$ has well-developed multi-peak  
structure which can perhaps be observed experimentally. Such dependence of  
the optical spectra on the twist angle is visible at low LL  
index, $n\lesssim 10$, and weak magnetic field, $B\lesssim 5$ Tesla.  
The strongest multi-peak structure should be observed at large inter-layer  
coupling, e.g. for the twist angle $38.2^0$.  
 
The work was supported by the Canada Research Chairs program.


\begin{thebibliography}{99} 
\bibitem[\ddag]{byline} Electronic address: 
tapash@physics.umanitoba.ca 
 
\bibitem{novo} 
A.K. Geim and K.S. Novoselov, Nat. Mater. {\bf 6}, 183 (2007). 
 
\bibitem{review} 
D.S.L. Abergel, V. Apalkov, J. Berashevich, K. Ziegler 
and T. Chakraborty, Adv. Phys. {\bf 59}, 261 (2010). 
 
\bibitem{mccann} 
E. McCann, Phys. Rev. B {\bf 74}, 161403 (2006). 
 
\bibitem{epitaxial} 
C. Berger, et al., Int. J. Nanotechnol. {\bf 7}, 383 (2010). 
 
\bibitem{reina} 
A. Reina, et al., Nano Lett. {\bf 9}, 30 (2009). 
 
\bibitem{walt_09} 
D.L. Miller, et al., Science {\bf 324}, 924 (2009). 
 
\bibitem{hass} 
J. Hass, et al., Phys. Rev. Lett. {\bf 100}, 125504 (2008). 
 
\bibitem{latil} 
S. Latil and L. Henrard, Phys. Rev. Lett. {\bf 97}, 036803 (2006). 
 
\bibitem{turbo} 
S. Shallcross, S. Sharma, and O.A. Pankratov, Phys. 
Rev. Lett. {\bf 101}, 056803 (2008). 
 
\bibitem{lopes07}  J.M.B. Lopes dos Santos, N.M.R. Peres, and A.H. Castro Neto, 
Phys. Rev. Lett. {\bf 99}, 256802 (2007). 
 
\bibitem{mele10} E.J. Mele, Phys. Rev B {\bf 81}, 161405 (2010). 
 
\bibitem{vadim_fqhe} 
V.M. Apalkov and T. Chakraborty, Phys. Rev. Lett. {\bf 97}, 126801 (2006). 
 
\bibitem{walt_10} 
Recent observation of the quantum Hall effects in epitaxial graphene 
seems to corroborate with our theoretical studies of pseudopotentials 
in a single layer graphene \protect\cite{vadim_fqhe}. See, Y.J. Song, et al., 
Nature {\bf 467}, 185 (2010). 
 
\bibitem{bilayer} V. Apalkov and T. Chakraborty, Phys. Rev. Lett. {\bf 105}, 
036801 (2010). 
 
\bibitem{haldane} F.D.M. Haldane, Phys. Rev. Lett. {\bf 51}, 605 (1983). 
 
\bibitem{fqhe_book} T. Chakraborty and P. Pietil\"ainen, {\it The 
Quantum Hall Effects} (Springer, Heidelberg, 1995), 2nd edition. 
 
\end{thebibliography}
\end{document}